\documentclass[twocolumn,fleqn]{svjour3}
\smartqed
\usepackage{graphicx}
\usepackage{caption}
\usepackage{amsmath}
\usepackage{amssymb}
\usepackage{xcolor}

\begin{document}

\title{Hierarchical organization of functional connectivity in the mouse brain: a complex network approach}
\author{Giampiero Bardella, Angelo Bifone, Andrea Gabrielli, Alessandro Gozzi, Tiziano Squartini}
\institute{Giampiero Bardella, Andrea Gabrielli, Tiziano Squartini$^*$\at
                Istituto dei Sistemi Complessi ISC-CNR, Universit\'a ``Sapienza'' di Roma, P.le A. Moro 5, 00185 Rome, Italy.
\\
                IMT Institute for Advanced Studies Lucca, P.zza S. Ponziano 6, 55100 Lucca, Italy.
\\
	    Alessandro Gozzi, Angelo Bifone\at
	    Italian Institute of Technology, Universit\'a di Trento, C.so Bettini 31, I-38068 Trento, Italy.
\\
              $^*$\email{tiziano.squartini@imtlucca.it}
}
\date{Received: date / Accepted: date}

\maketitle

\begin{abstract}
This paper represents a contribution to the study of the brain functional connectivity from the perspective of complex networks theory. More specifically, we apply graph theoretical analyses to provide evidence of the modular structure of the mouse brain and to shed light on its hierarchical organization. We propose a novel percolation analysis and we apply our approach to the analysis of a resting-state functional MRI data set from 41 mice. This approach reveals a robust hierarchical structure of modules persistent across different subjects. Importantly, we test this approach against a statistical benchmark (or null model) which constrains only the distributions of empirical correlations. Our results unambiguously show that the hierarchical character of the mouse brain modular structure is not trivially encoded into this lower-order constraint. Finally, we investigate the modular structure of the mouse brain by computing the Minimal Spanning Forest, a technique that identifies subnetworks characterized by the strongest internal correlations. This approach represents a faster alternative to other community detection methods and provides a means to rank modules on the basis of the strength of their internal edges.
\end{abstract}

\section{Introduction}

The brain can be represented as a network of connected elements at different spatial scales, from individual neurons to macroscopic, functionally specialized structures \cite{martijn,yao,bullmore,sporns,bifons}. Interestingly, neuroimaging data, like those obtained with functional Magnetic Resonance Imaging (fMRI) techniques, naturally lend themselves to a network representation, thus attracting the interest of both graph-theorists and network scientists towards a study of the topological properties of brain connectivity structures \cite{wang}. Indeed, correlations between fMRI signals arising from responses to stimuli or from spontaneous fluctuations in the brain resting-state can be interpreted as a measure of functional connectivity between remote brain regions and represented as edges in a graph. Moreover, alterations in the strength and structure of functional connectivity networks have been observed in groups of patients suffering from several brain diseases, including Alzheimer, Autism and Schizophrenia, thus providing potential markers of neuropsychiatric illness \cite{martijn,liska,biswal,rosazza,zhang,fox}.

Of particular interest is the study of the modular structure of these networks, i.e. the presence of clusters of nodes that are more tightly connected among themselves than with nodes in other network substructures \cite{martijn,yao,lambio,achab,lazaros}. A modular structure has been observed for different types of brain networks (functional and structural) and in different species, including humans, primates and rodents \cite{liska,achab,lazaros,bifone}. Functional connectivity networks derived from fMRI experiments in human subjects exhibit a hierarchical structure of modules-within-modules \cite{bullmore,sporns}. It has been suggested that hierarchical modularity may confer important evolutionary and adaptive advantages to the human brain by providing intermediate modules that can respond to the evolutionary or environmental pressure without jeopardizing the function of the entire system \cite{simon}. A similar hierarchical organization has been observed in other species, e.g. non-human primates, but not in lower species, like the worm \emph{C. Elegans}, which seems to have a modular network of neurons that is not hierarchically organized \cite{worm,worm2}. Here, we investigate the modular structure of the mouse brain and its hierarchical organization using a graph theoretical approach.

Percolation analysis, a tool derived from statistical physics, provides a powerful means to investigate the hierarchical organization of networks \cite{barabasi,null}. This approach is based on the assessment of the fragmentation of a network as weaker edges are gradually removed from the graph. A striking demonstration of this hierarchical organization is the presence of multiple percolation thresholds \cite{lazaros}, whereby disaggregation of modules occurs abruptly for critical values of the control parameter, $p_c$. On the contrary, application of this analysis to Erd\"{o}s-Renyi random graphs \cite{barabasi,bollobas,null} shows a single threshold value, separating two phases characterized by different topological features. Below the threshold (i.e. for $p<p_c$) several tree-like components are observed whose size is of the order of $\ln N$ - with $N$ the total number of nodes. Above the threshold (i.e. for $p>p_c$), instead, a single giant component appears, whose structure admits cycles and from which tree-like structures (whose size is again of the order of $\ln N$) are excluded \cite{barabasi,bollobas}.

\begin{figure}[t!]
\centering
\hspace{1.5mm}\includegraphics[scale=0.381,angle=-90]{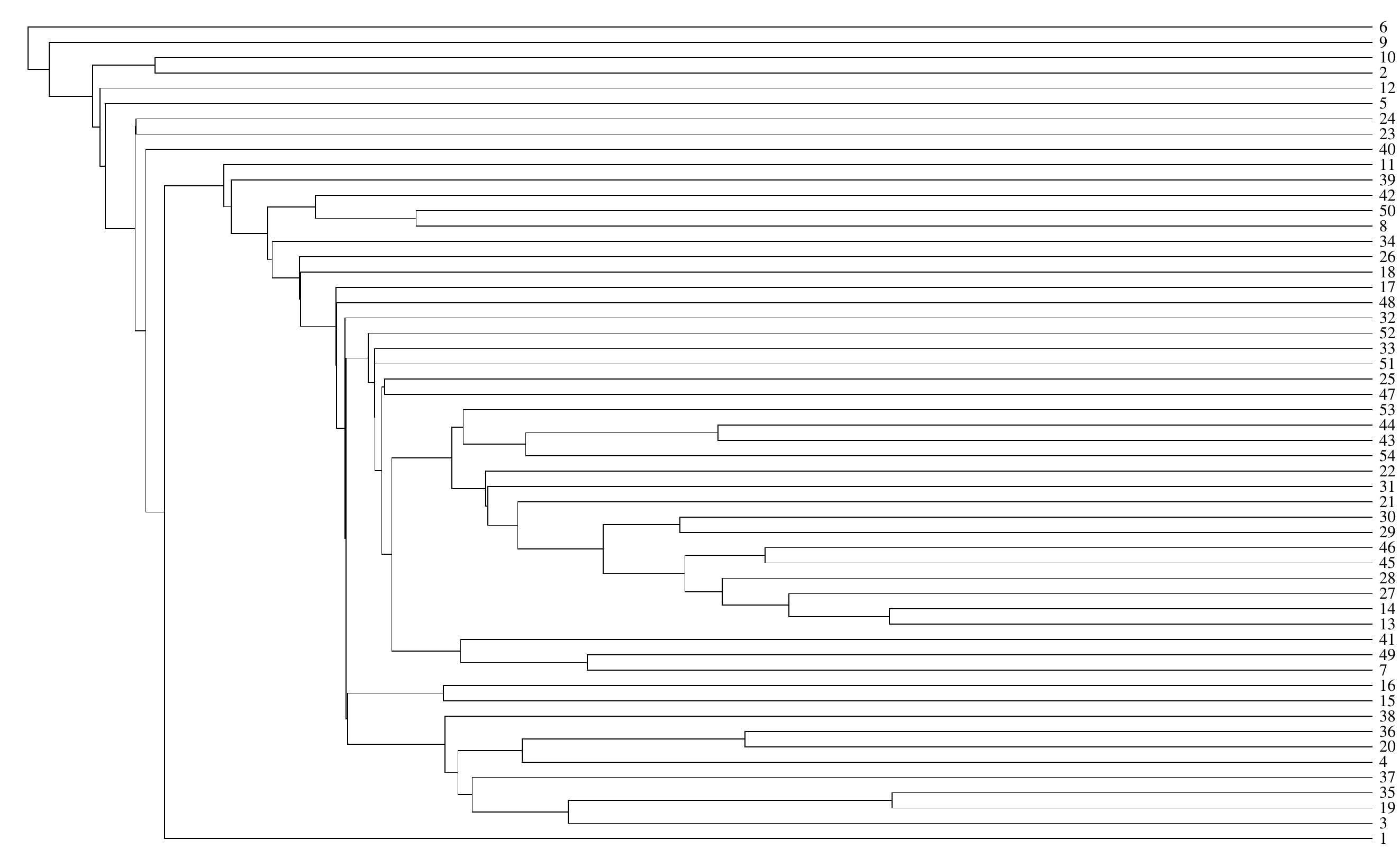}\\
\includegraphics[width=0.45\textwidth,angle=-90]{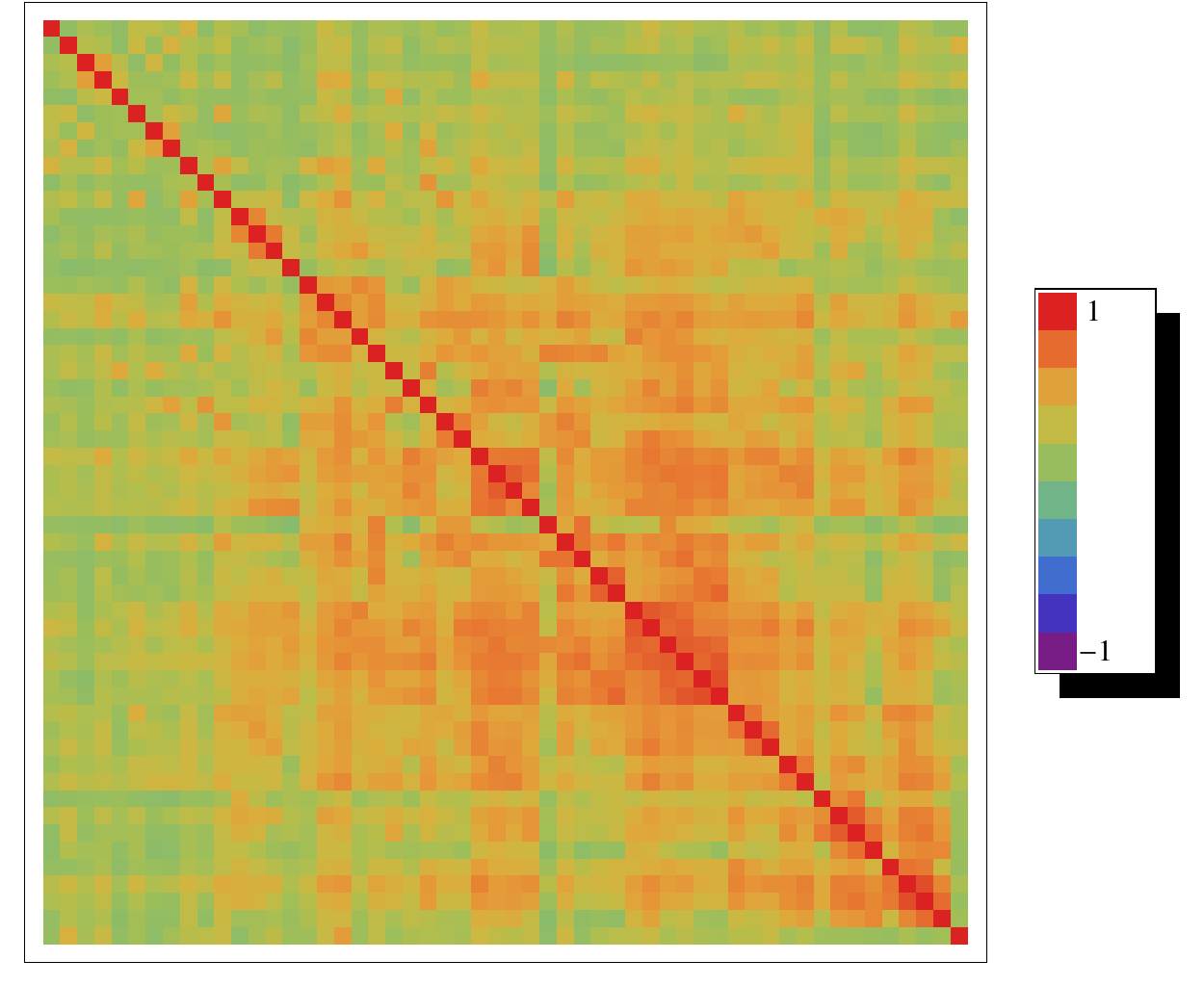}
\caption{Dendrogram and correlation matrix for the average brain, induced by the dissimilarity measure $D_{ij}=1-C_{ij},\:\forall\:i, j$.}
\label{fig1}
\end{figure}

Here we have analyzed functional connectivity networks constructed from a large resting state fMRI dataset from mice to assess the presence of multiple percolation thresholds. Specifically, we have applied standard percolation analysis and variations thereof to assess the hierarchical modular structure in this species. Importantly, we have applied novel approaches to avoid some of the pitfalls that may affect more conventional analysis of functional connectivity networks. Indeed, it will be shown that traditional percolation detects a modular structure even in random networks, thus making it necessary to introduce a null model in order to correctly asses the statistical significance of the percolation analysis. Here we introduce a novel null model, independent of the choice of a particular threshold and resting exclusively on the information encoded into the correlation matrix. Moreover, we propose the use of an algorithm to calculate the closest correlation matrix to a given symmetric matrix, thus ensuring that the proposed null model has the peculiar features of a proper correlation matrix.

We have complemented our percolation analysis by computing the Minimal Spanning Forest (MSF). Althought the MSF is not, by itself, a community detetction technique, it represents a faster alternative for the identification of modules, defined by the strength of the functional relations between nodes. Such modules can be, in turn, linked to obtain the Minimal Spanning Tree (MST), which provides the ``backbone'' of the mouse brain functional connectivity.

These methodological developments make it possible to assess the presence of a hierarchically-organized modular structure in the mouse brain, both at the level of population and of individual subjects.

\begin{figure}[t!]
\centering
\includegraphics[width=0.45\textwidth]{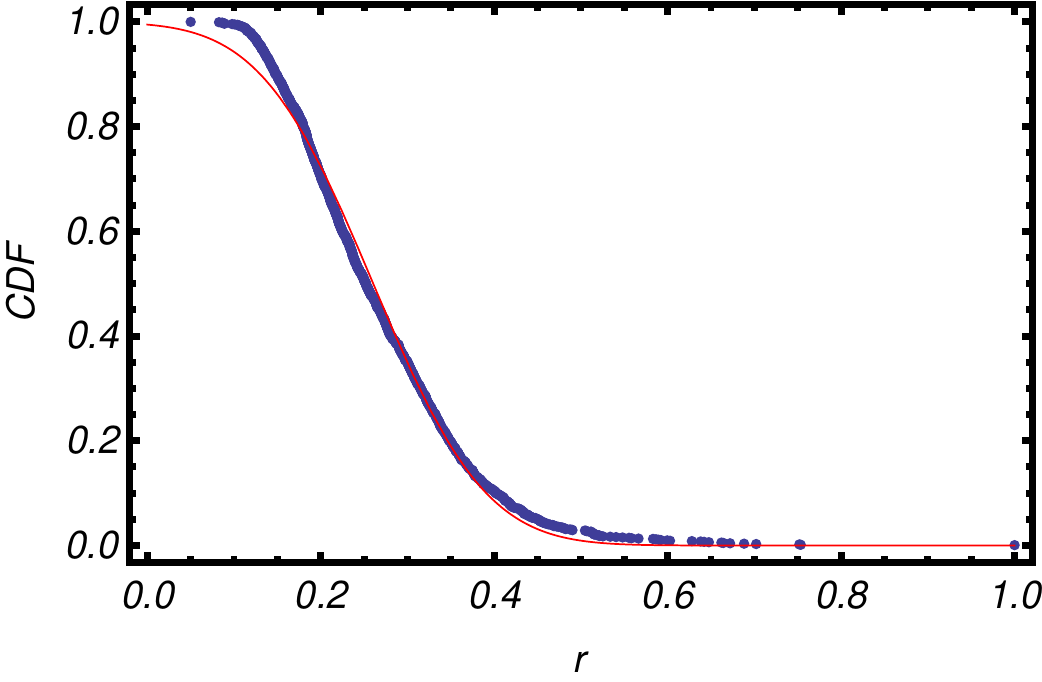}
\caption{Empirical CDF of the correlations for the average brain (blue trend) and CDF of a gaussian distribution whose means and standard deviations have been estimated through the maximum-of-the-likelihood procedure (red trend).}
\label{fig2}
\end{figure}

\section{Methods}

\begin{figure}[t!]
\centering
\includegraphics[width=0.467\textwidth]{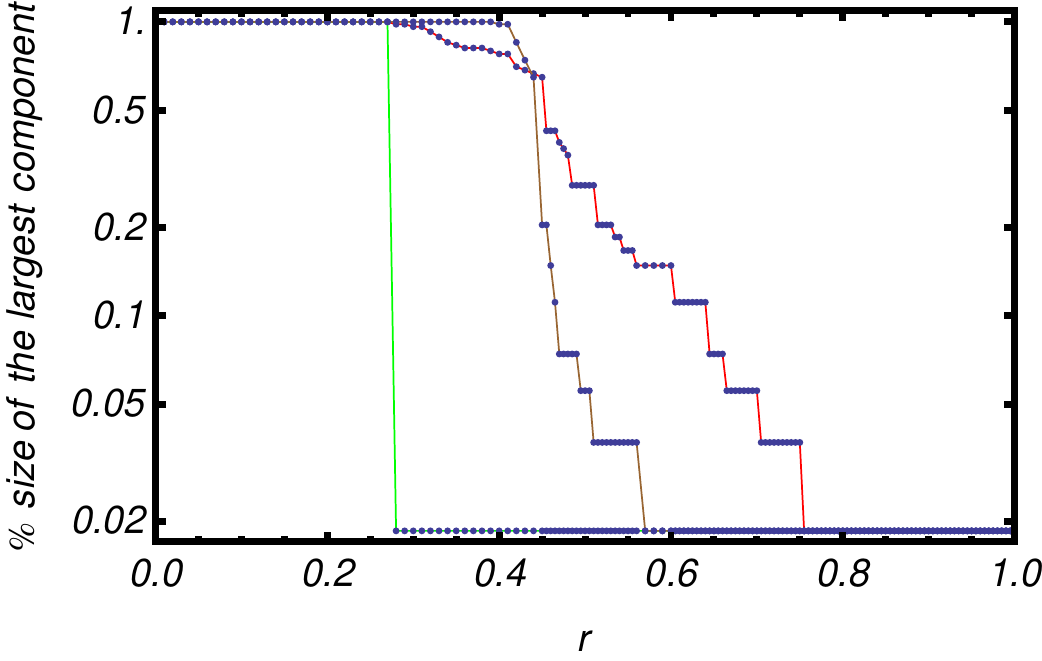}\\
\hspace{3mm}\includegraphics[width=0.45\textwidth]{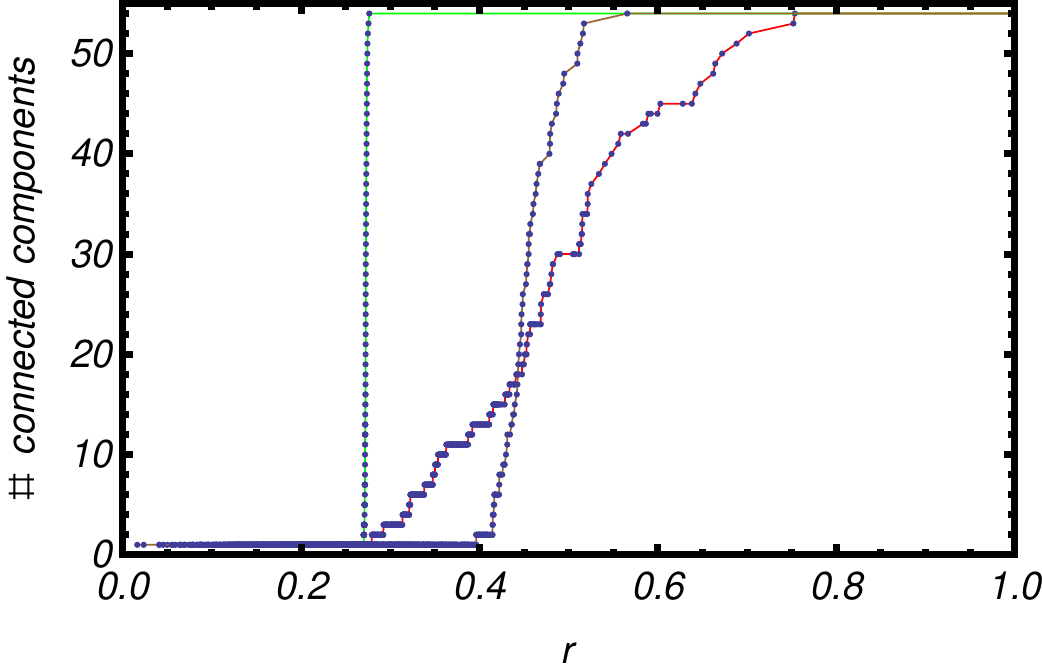}
\caption{Comparison between the usual percolation analysis (top panel) and our modified percolation analysis (bottom panel) run on the average brain (red trend), on a randomized version of it, retaining the same empirical distribution of correlations (brown trend) and on the ensemble-averaged matrix (green trend). While the usual percolation analysis detects a hierarchical modular structure even on the null model, thus making it difficult to asses the statistical significance of the observed patterns, our modified percolation analysis enables discrimination between the real and the random cases.}
\label{fig3}
\end{figure}

\subsection{Data acquisition and data pre-processing}

The data-set used for this analysis has been reported in a recent paper \cite{liska,sfora}, where experimental details are extensively described. In short, MRI experiments were performed on male 20-24 week old C57BL/6J (B6) mice ($n=41$, Charles River, Como, Italy). Mice were anaesthetised with isoflurane (5\% induction), intubated and artificially ventilated under 2\% isoflurane maintenance anesthesia. All experiments were performed with a 7.0 T MRI scanner (Bruker Biospin, Milan) using an echo planar imaging (EPI) sequence with the following parameters: TR/TE 1200/15 ms, flip angle 30 degrees, matrix $100\times100$, field of view $2\times2$ cm$^2$, 24 coronal slices, slice thickness 0.50 mm, 300 volumes and a total rsfMRI acquisition time of 6 minutes. All experiments were conducted in accordance with the Italian law (DL 116, 1992 Ministero della Sanit\'a, Roma) and the recommendations in the ``Guide for the Care and Use of Laboratory Animals'' of the National Institutes of Health. Animal research protocols were also reviewed and consented to by the animal care committee of the Istituto Italiano di Tecnologia (permit 07-2012). All surgical procedures were performed under anesthesia.

The mouse brain was parcellated into 54 macro-regions (27 per hemisphere) described in the Appendix. Resting state fMRI signals from individual image voxels were averaged across each region of interest (ROI) to generate 54 time-series of approximately 300 s duration. The 54 collected time-series were pairwise correlated calculating the Pearson coefficient and organized in a $54\times54$ symmetric matrix describing the resting-state connectivity network for each mouse.

Image preprocessing was carried out using tools from FMRIB Software Library (FSL, v5.0.6 \cite{wtw1,jenk}) and AFNI (v2011\_12\_21\_1014 \cite{wtw2}). RsfMRI time series were despiked (AFNI/3dDespike), corrected for motion (AFNI/3dvolreg) and spatially normalised to an in-house C57Bl/6J mouse brain template \cite{sfora2} (FSL/FLIRT,12 degrees of freedom). The normalised data had a spatial resolution of $0.2\times0.2\times0.5$ mm$^3$ ($99\times99\times24$ matrix). Head motion traces and mean ventricular signal (averaged fMRI time course within a manually-drawn ventricle mask) were regressed out of each of the timeseries (AFNI/3dDeconvolve). To assess theeffectof global signal removal, separate rsfMRI time series with the whole-brain average time course regressed out were also generated. All rsfMRI time series were spatially smoothed (AFNI/3dmerge, Gaussian kernel of full width at half maximum of 0.5 mm) and band-pass filtered to a frequency window of 0.01-0.08 Hz (AFNI/3dBandpass) \cite{sfora2}.

In order to create an average adjacency matrix describing brain functional connectivity at the population level, subject-wise matrices were first Fisher-transformed, averaged across subjects and then back-transformed.

\subsection{Percolation analysis}

The percolation analysis proposed by Makse et al. \cite{lazaros} includes the following steps: {\it a)} a threshold parameter $p$, ranging between 0 and 1 (and thus interpretable as a probability), is chosen; {\it b)} the links corresponding to the correlations below the threshold are removed and the size of the giant component $C$ (i.e. the largest connected component) is computed; {\it c)} the parameter $p$ is varied and $C$ is evaluated for different thresholds.

This procedure ignores the complex evolution of the structure of the whole network, which is not captured by the giant component only. This becomes a relevant issue when the classical percolation is applied to small networks, i.e. to networks for which no giant component is clearly distinguishable: in this case the signal provided by this kind of analysis may be rather noisy, thus misrepresenting the modular structure of the brain at the global level.

\begin{figure}[t!]
\centering
\hspace{3mm}\includegraphics[width=0.435\textwidth]{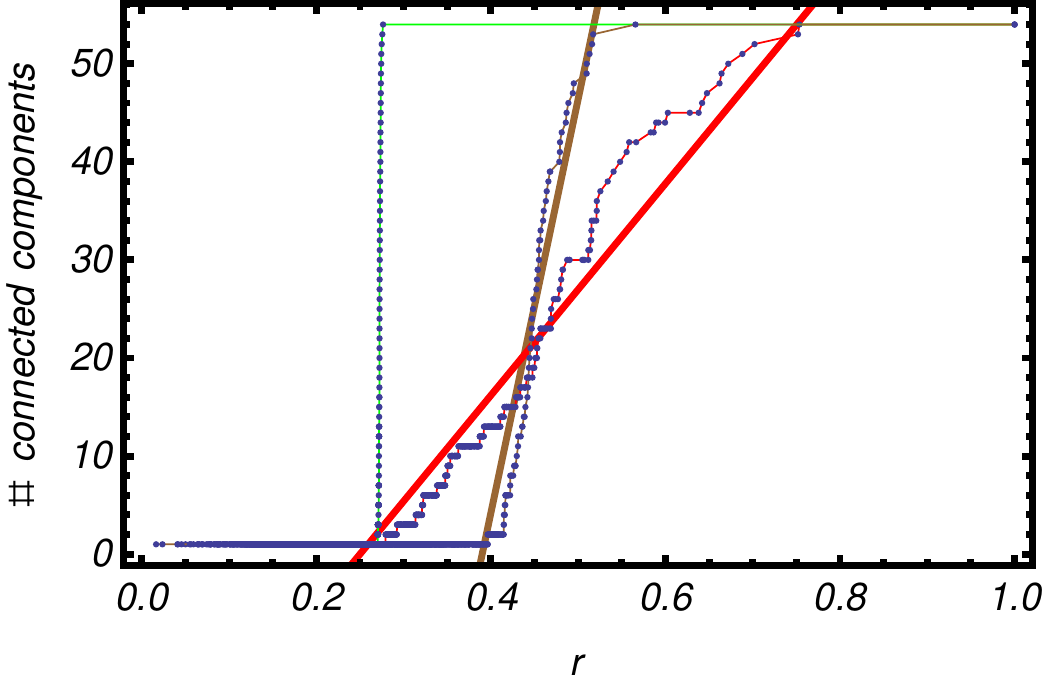}\\
\vspace{5mm}
\includegraphics[width=0.45\textwidth]{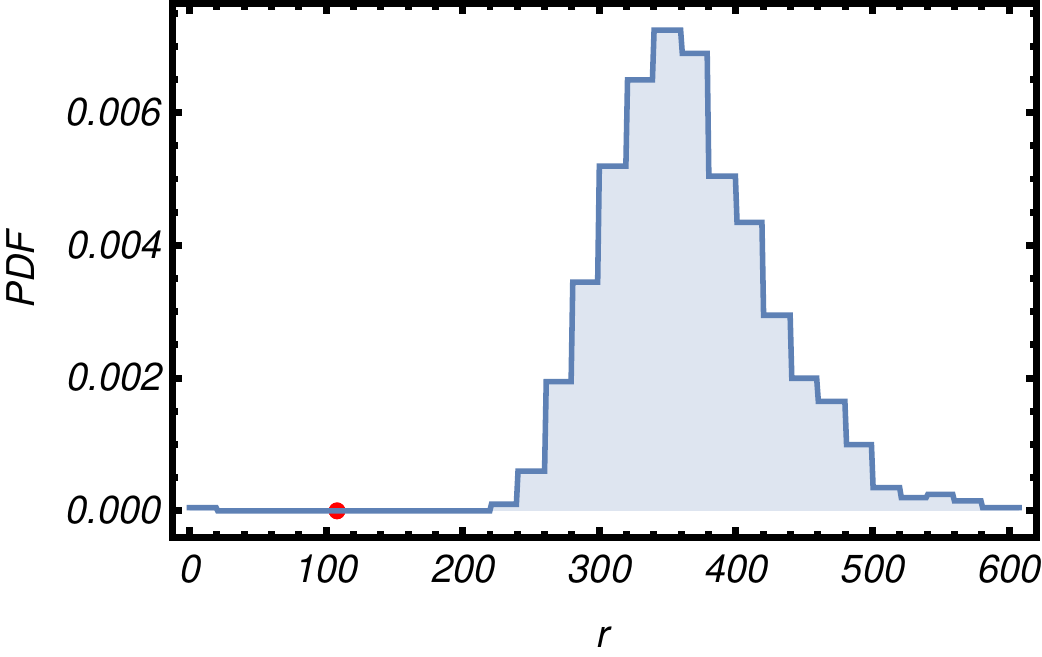}
\caption{In order to assess the statistical significance of the results of our modified percolation analysis, a test is needed. Top panel represents the test statistics we have chosen: the slope of the percolation plot of both the average brain (red trend) and of a randomized version of it, retaining the same empirical distribution of correlations (brown trend). Bottom panel: ensemble distribution of our test statistics; the red point represents the (statistically significant) observed value of the latter.}
\label{fig4}
\end{figure}

In order to overcome this drawback, we propose a variation of the percolation analysis along the following lines: {\it a)} all experimentally determined correlation coefficients are listed in increasing order; {\it b)} starting from the lowest value, each entry in the list is chosen as a threshold; {\it c)} all the links corresponding to the correlations below the threshold are removed; {\it d)} the number of connected components characterizing the remaining part of the network is computed. 

Beside providing a much more precise picture of the dynamics of the brain at the global level, our variation of the percolation analysis is also more robust, since our signal results from the fragmentation of many different components at the same time and is thus less prone to the statistical noise which, instead, accompanies the fragmentation of the giant component only.

Moreover, while each step of the classical percolation analysis is always mappable into a step of our method, the reverse is not true: the detection of a newly disconnected module from a secondary component would be missed by the classical percolation analysis (which focuses on the giant component only).

\subsection{A statistical benchmark for mice brains}

In order to define to what extent the stepwise structure highlighted by the percolation analysis is significant, we need to compare the results with a proper statistical benchmark. In other words, in order to understand whether the ``stepwise behavior'' is a mere consequence of lower- order constraints or a genuine sign of self-organization we need to define a proper null model.

As a first step, we have calculated the empirical probability distributions of the entries of the correlation matrix characterizing each subject in our sample and fitted them to normal distributions, whose means and standard deviations were estimated through the maximum-of-the-likelihood procedure. We have also repeated this analysis for the average mouse, i.e. the brain functional connectivity at the population level. In all cases, the distributions of the elements of the correlation matrices appeared to be well behaved, with nearly Gaussian distributions.

\begin{figure*}[t!]
\centering
\includegraphics[width=0.26\textwidth]{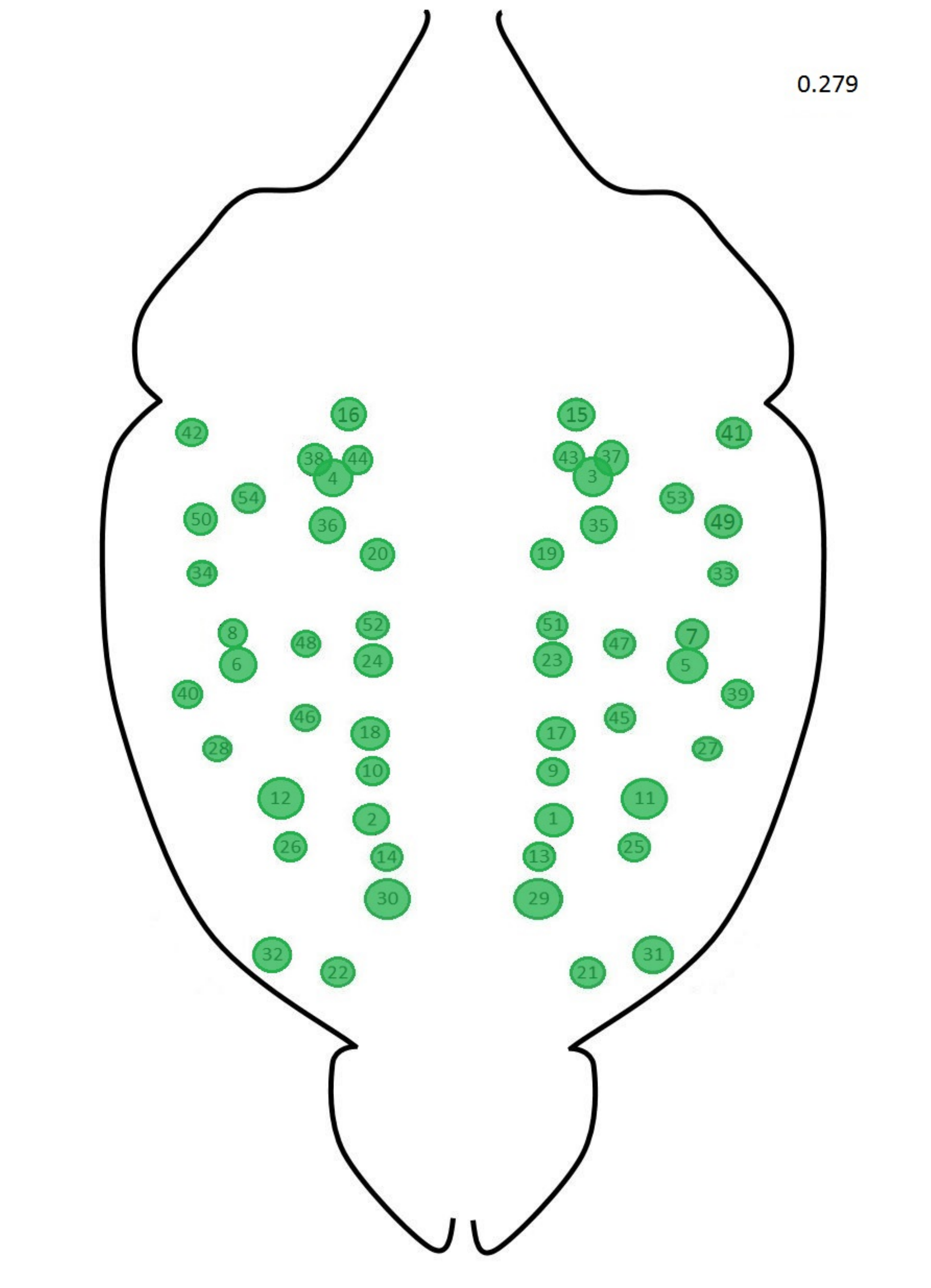}
\includegraphics[width=0.26\textwidth]{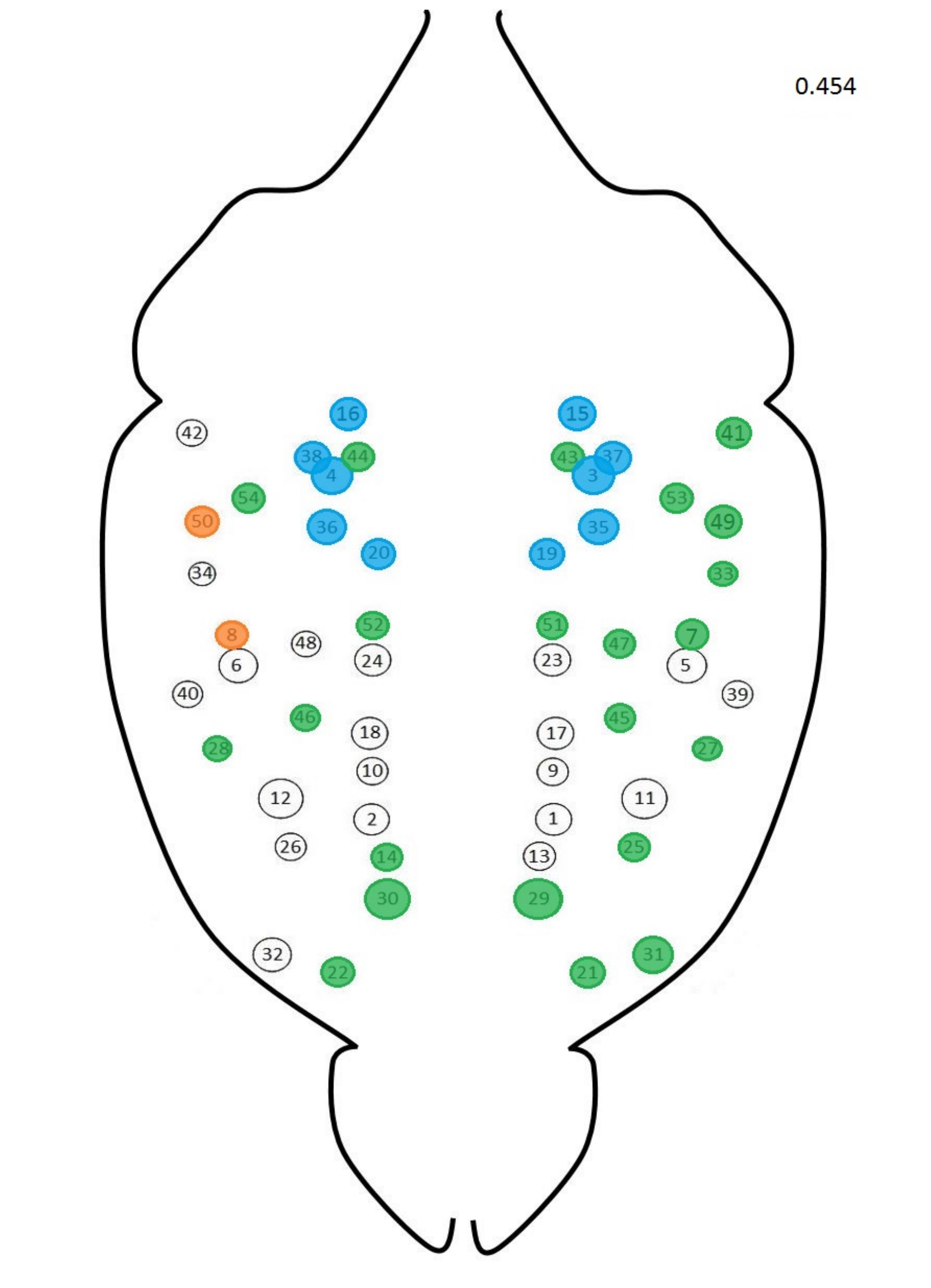}
\includegraphics[width=0.26\textwidth]{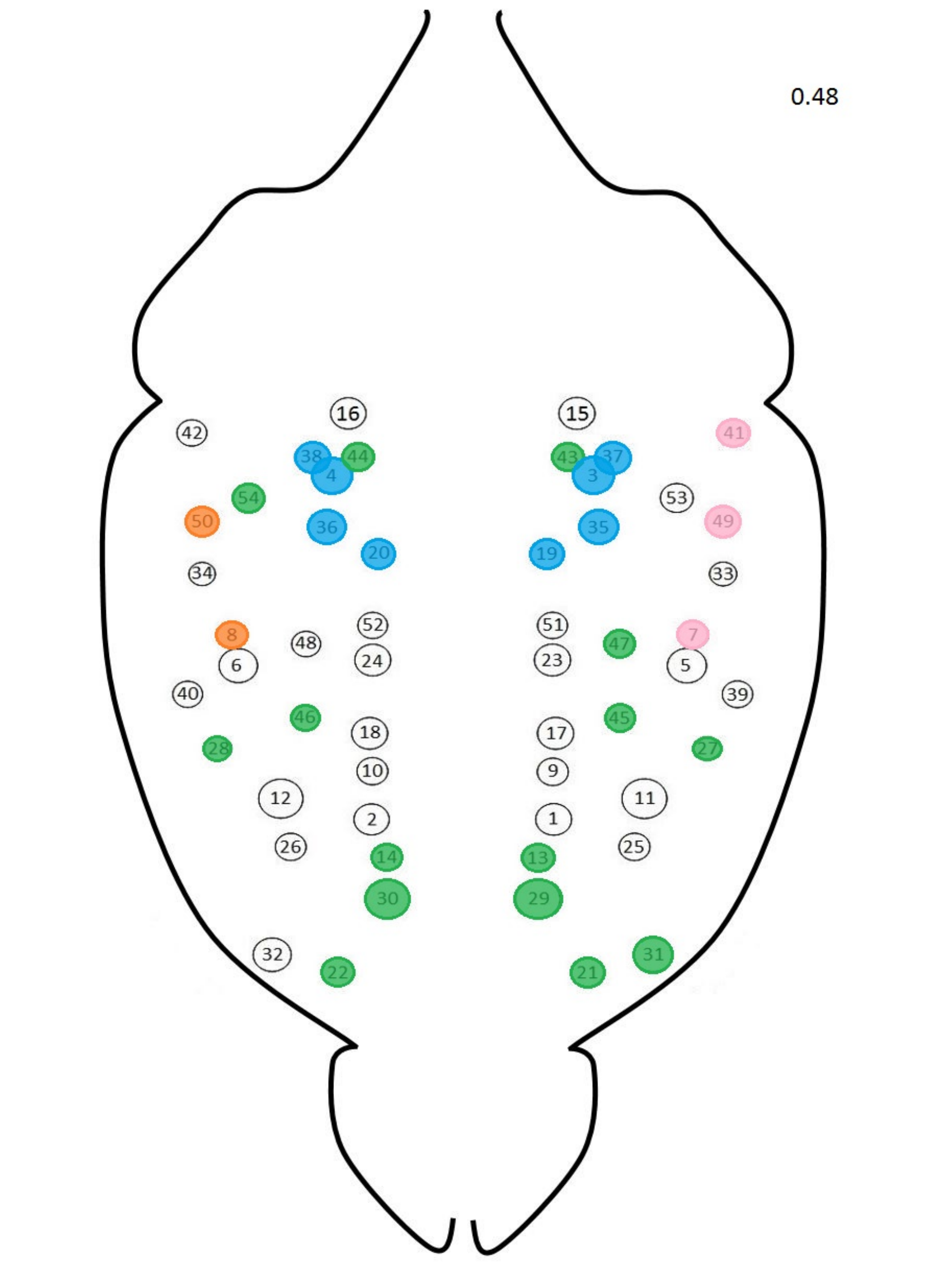}\\
\includegraphics[width=0.26\textwidth]{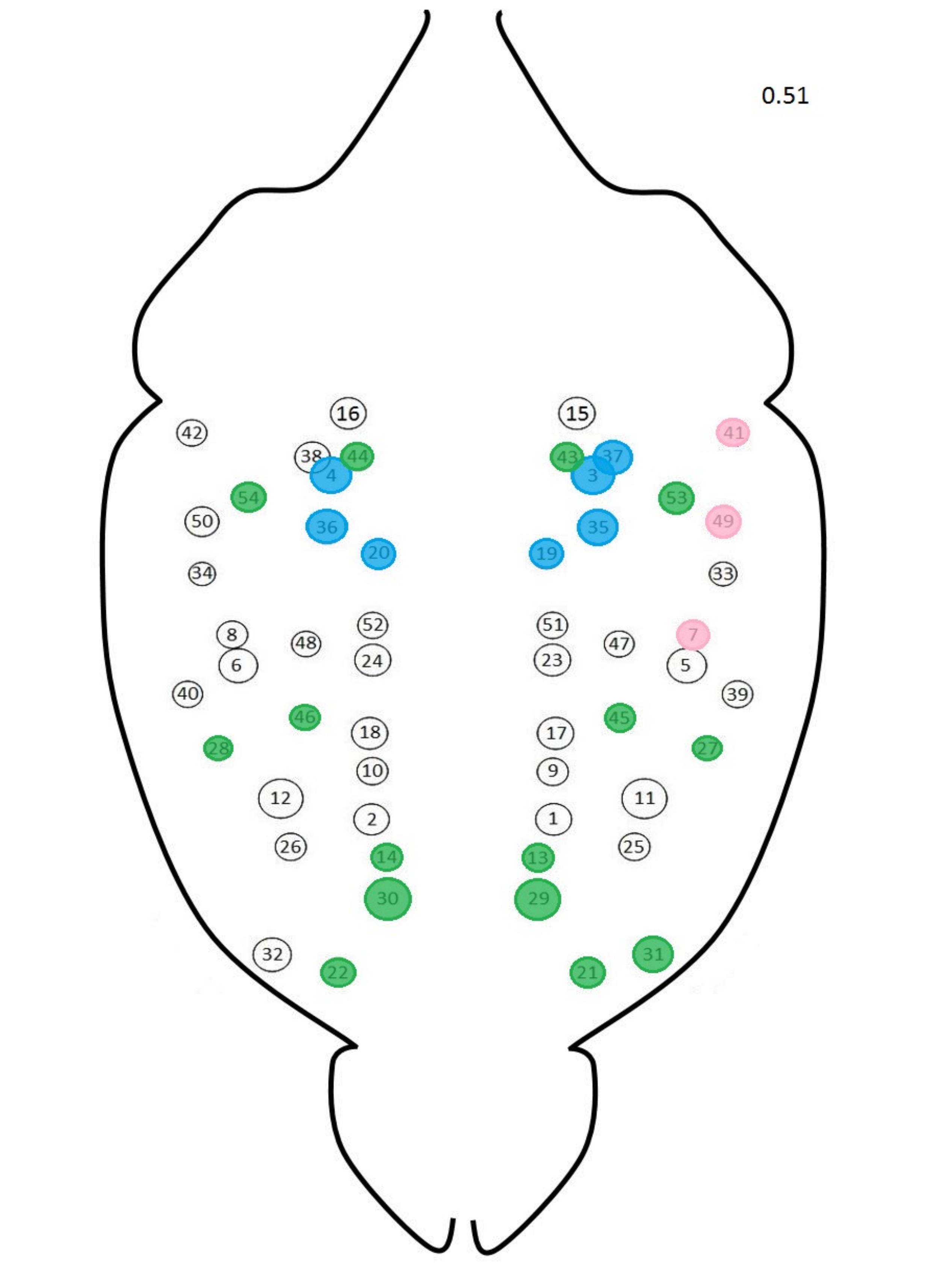}
\includegraphics[width=0.26\textwidth]{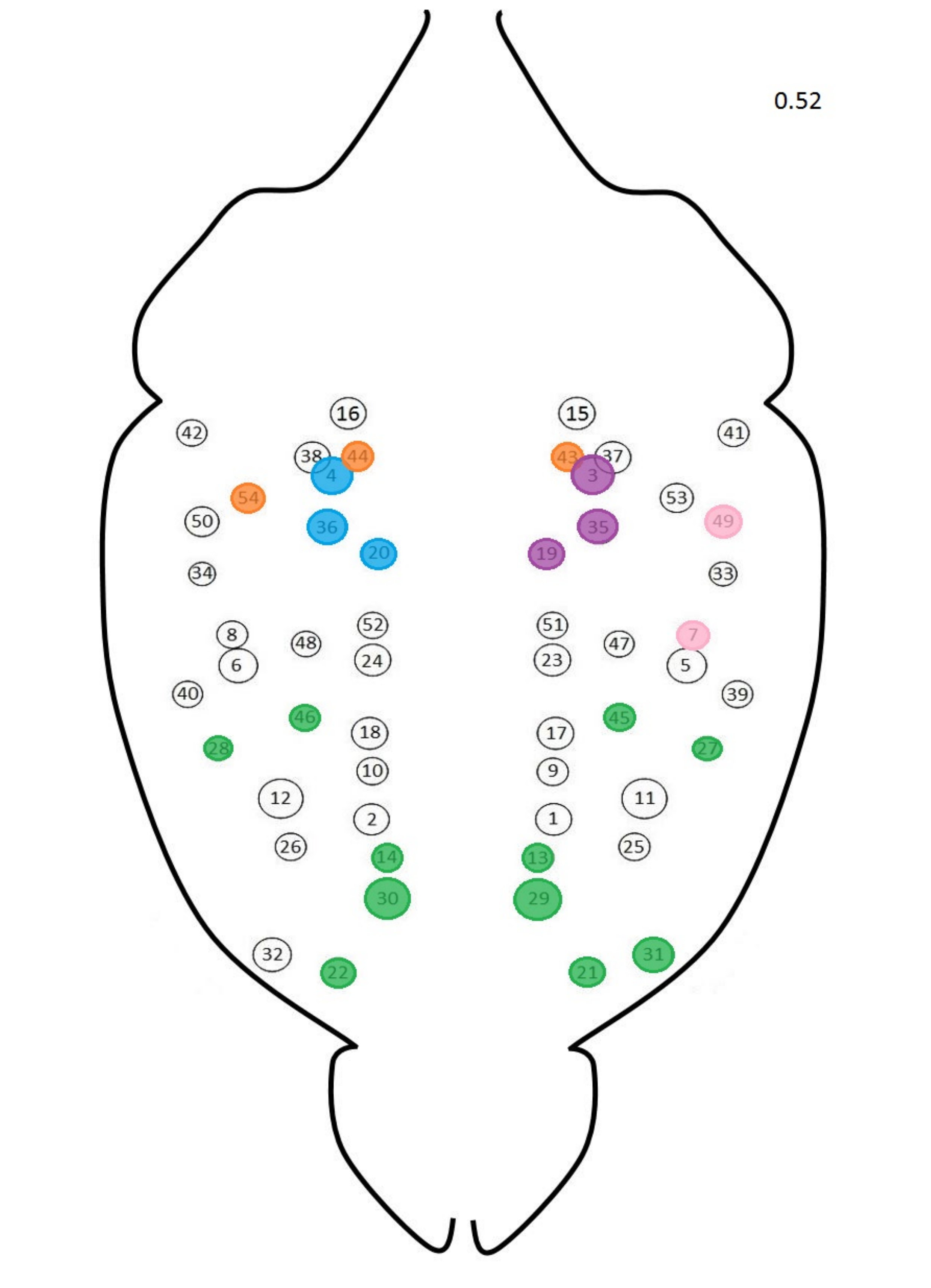}
\includegraphics[width=0.26\textwidth]{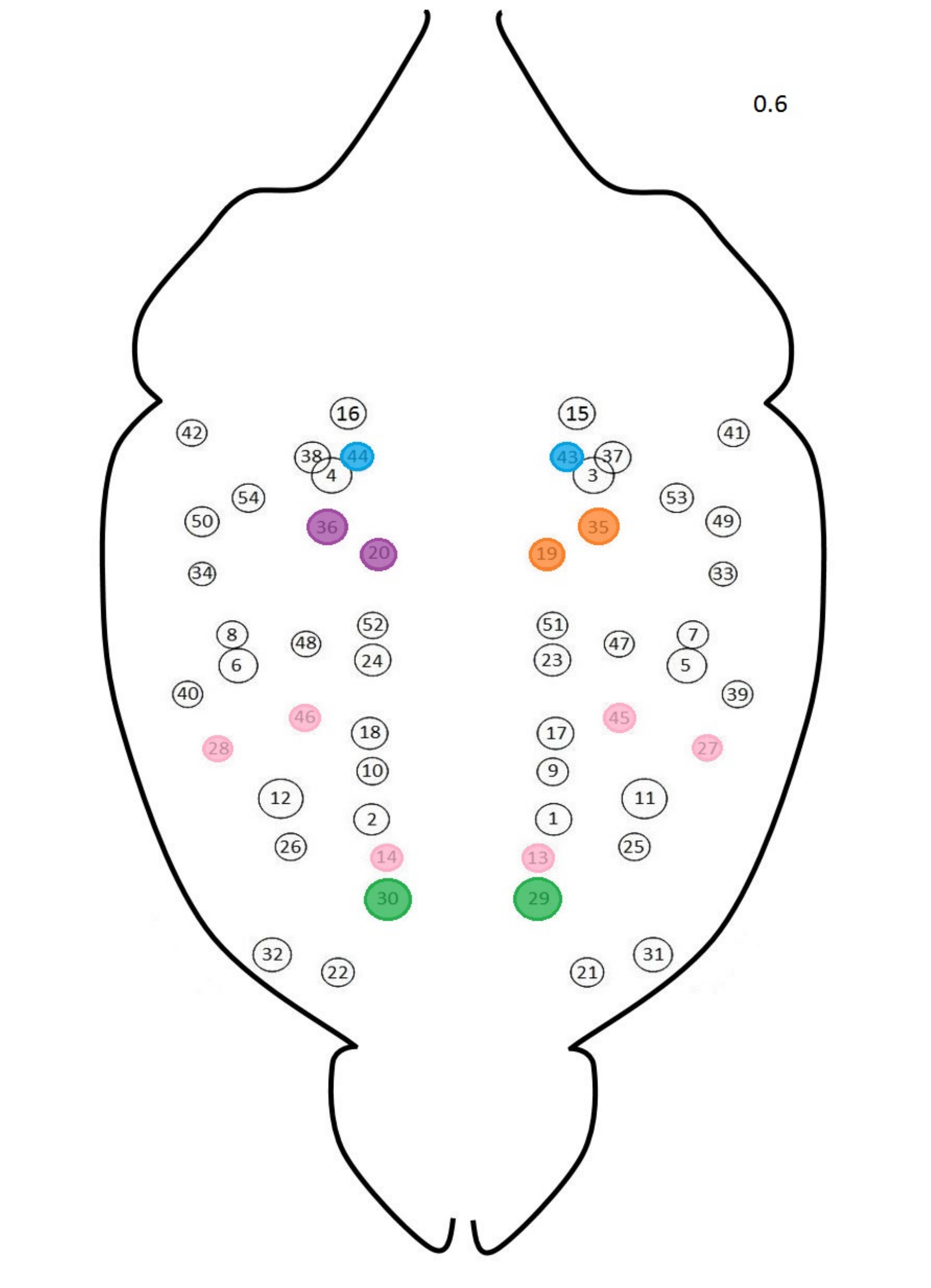}\\
\includegraphics[width=0.26\textwidth]{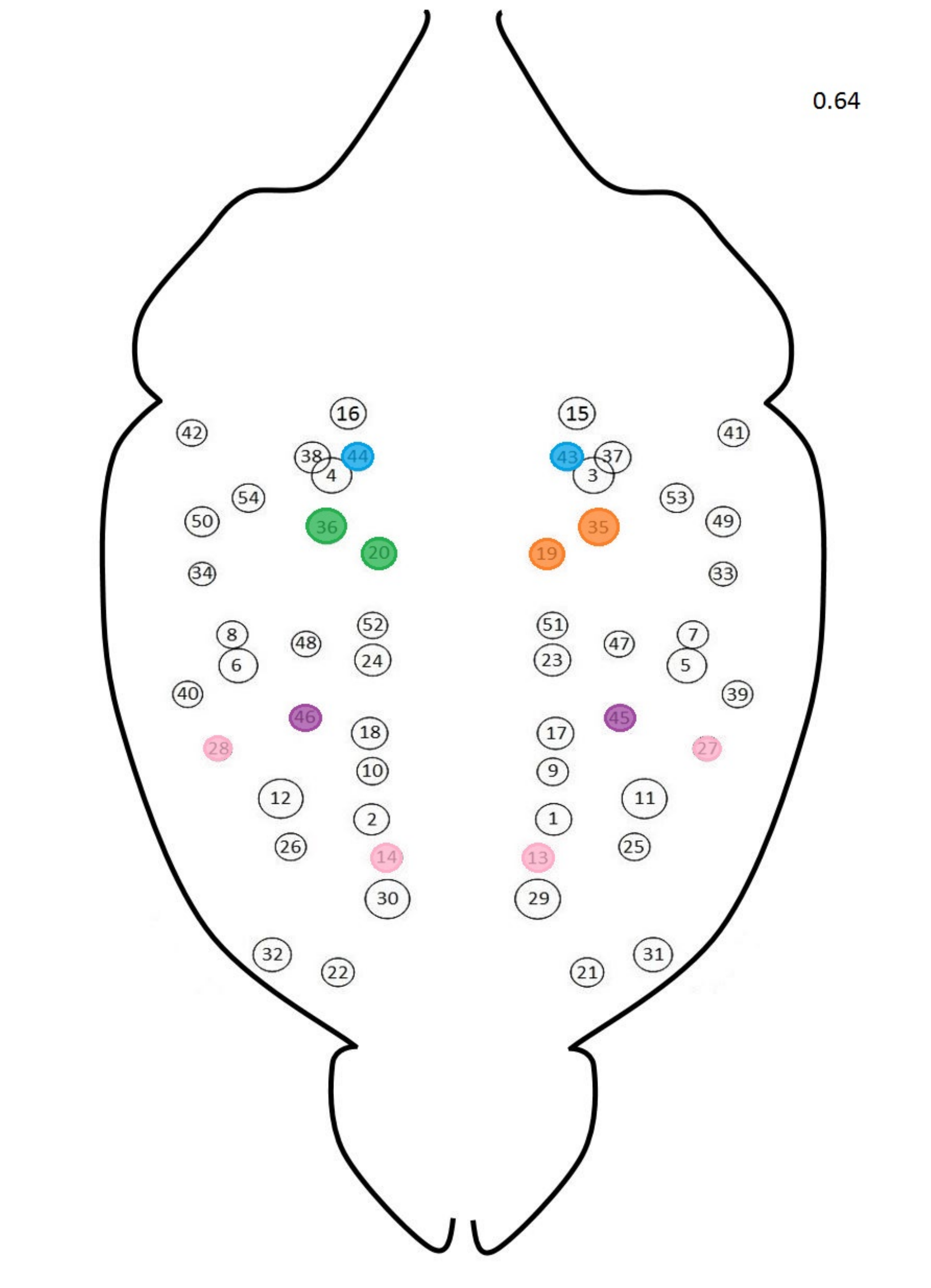}
\includegraphics[width=0.26\textwidth]{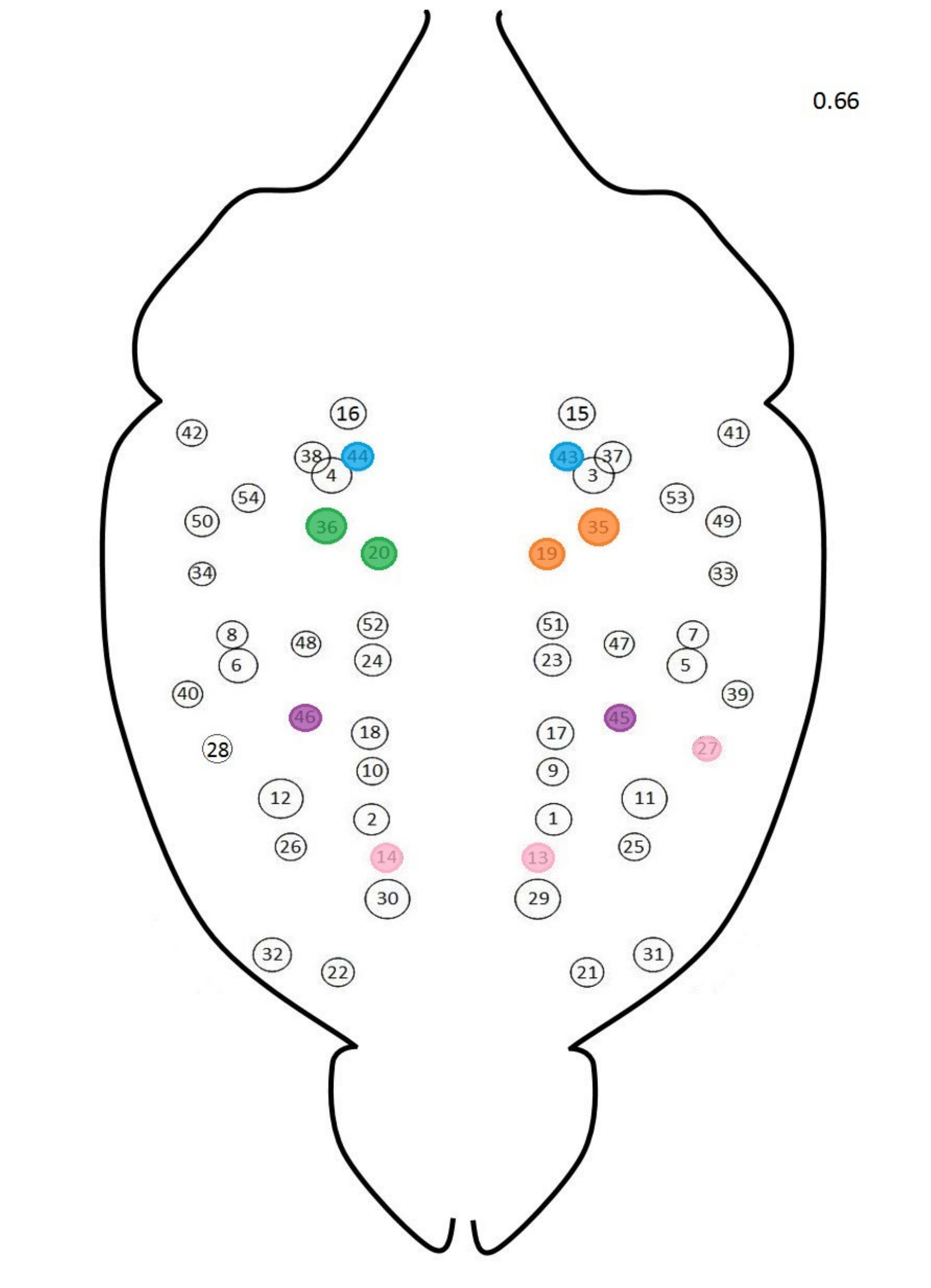}
\includegraphics[width=0.26\textwidth]{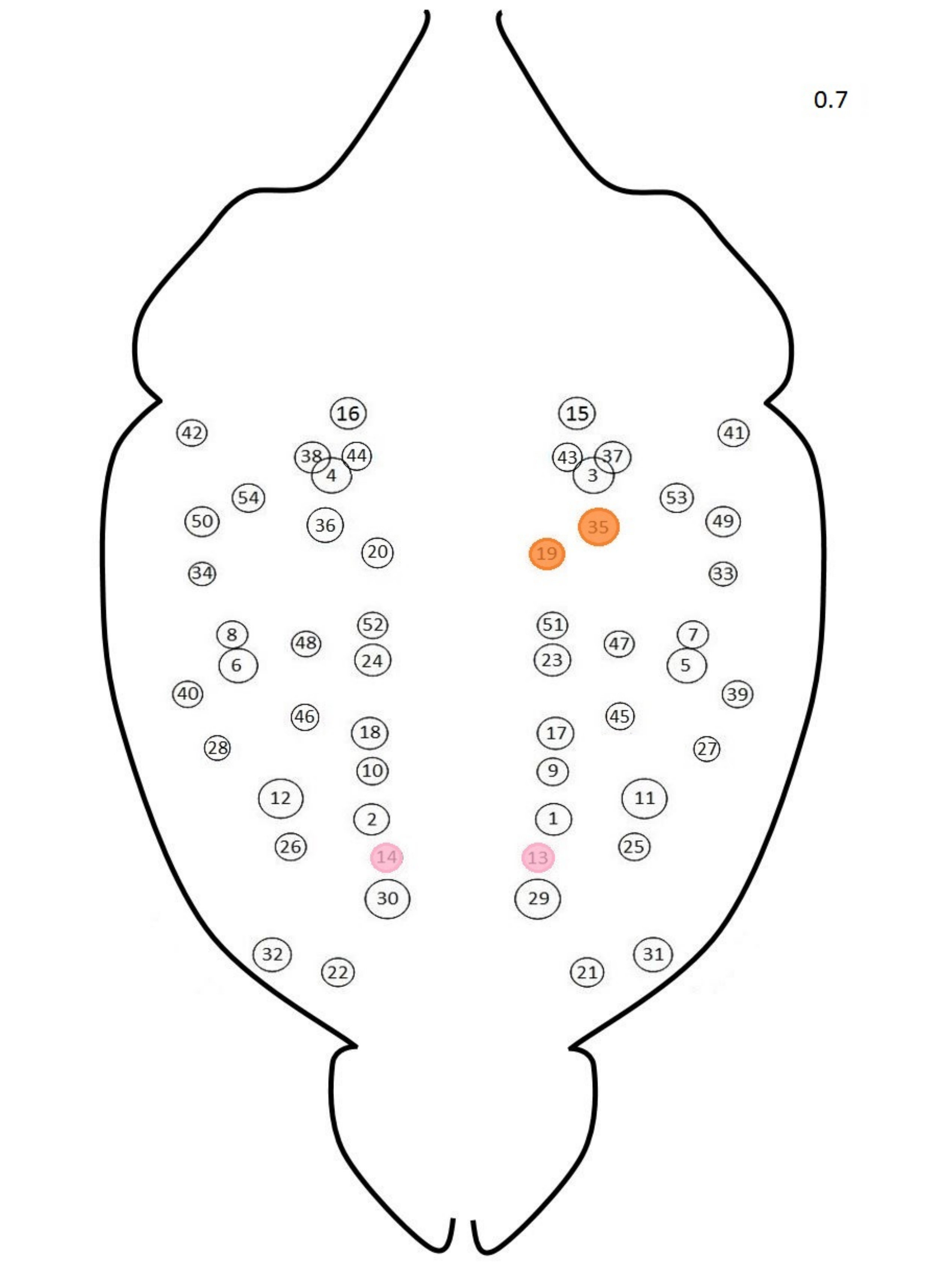}
\caption{Each group of areas detected by our percolation analysis in correspondence of a given correlation value is composed by many sub-modules, whose presence is evidenced by rising the threshold value. A clear example is provided by the blue area detected for $r_{th}=0.51$, comprising the anterio-dorsal hippocampus, the dentate gyrus and the posterior dentate gyrus - i.e. areas 3, 4, 19, 20, 35, 36. Upon rising the threshold to $r_{th}=0.52$, two subgroups appear, composed respectively by the right and left parts - i.e. 3, 19, 35 and 4, 20, 36 - of the aforementioned areas (evidenced in blue and purple). Further rising the threshold to $r_{th}=0.6$, the two subgroups reveal a core structure defined by the pairs 19, 35 and 20, 36. This finding confirms the hierarchical character of the mouse brain modular structure. See also the map of the neuroanatomical ROI in Appendix.}
\label{fig5}
\end{figure*}

Secondly, we have generated a ``null brain'', by drawing correlations from the corresponding normal distributions. However, this procedure does not guarantee that true correlation matrices, which should be positive-definite, are obtained: in fact, although the synthetic matrices can be chosen to be sym-metric and with unitary elements on the main diagonal, they may still have negative eigenvalues. This problem can be solved by implementing the procedure illustrated in \cite{higham} where a fast algorithm for comput- ing the nearest correlation matrix to a given, symmetric, one is described. The last step of our method consists in the implementation of this procedure.

\section{Results}

\subsection{Average correlation matrix}

We first focus on the average correlation matrix, defined by the sample mean (i.e. over all individuals) of each back-transformed pair-specific correlation coefficient. Fig. \ref{fig1} shows the average correlation matrix whose rows and columns have been reordered according to the \emph{dissimilarity measure}

\begin{equation}
D_{ij}=1-C_{ij},\:\forall\:i, j.
\label{dissp}
\end{equation}

The algorithm we have adopted proceeds by computing, at each step, the minimum dissimilarity between pairs of areas and clustering them together. In other words, clusters are grouped according to the minimum intercluster dissimilarity, a linkage rule also known as ``single-linkage'' clustering \cite{single}. The same algorithm can be used to generate the corresponding dendrogram.

While negative correlations are pronounced in subject-wise matrices they tend to be averaged-out in the population-wise matrix, whose terms are all positive. Although this confirms the larger inter-subject variability of negative correlations with respect to the positive ones, it affects the nested structure of the average matrix, which is far less pronounced than for the single individuals: nonetheless, nested red square-shaped patterns along the diagonal are still clearly visible.

The distribution of edge values of the average matrix is shown in fig. \ref{fig2}, alongside with the normal distribution whose mean and variance have been estimated through the maximum-of-the-likelihood procedure. The deviation of the distribution of experimentally-determined correlations form the normal distribution is larger for this matrix than for the individual ones.

\paragraph{Percolation analysis.}

The results of the classical and modified percolation analyses are shown in fig. \ref{fig3}. 

Our method identifies multiple steps for increasing threshold, corresponding to the stable partitions of the network \cite{martijn,lazaros} highlighted in fig. \ref{fig5}. The plateaus indicate the presence of connections whose removal does not affect the number of connected components, indicating that these links are not critical in determining the structure of functional correlations.

Fig. \ref{fig5}, shows that each connected group of areas detected in correspondence of a given correlation value is composed by many nested modules, whose hierarchical organization emerges form the application of  higher thresholds. Two main groups of areas can be clearly identified (colored in blue and green in fig. \ref{fig5} and detected for $r_{th}\simeq0.45$). The first group (colored in green) regions include the cingulate cortex, the motor cortex, the medial prefrontal cortex and the primary somatosensory cortex \cite{martijn,liska,rosazza}. The second group (colored in blue) is constituted by areas 3, 4, 19, 20, 35 and 36 (i.e. anterio-dorsal hippocampus, the right dentate gyrus and the right posterior gyrus), all parts of the hippocampal formation.

Upon rising the threshold to $r_{th}=0.52$, sub-areas appear: for example, the hippocampus splits into right and left part - i.e. 3, 19, 35 and 4, 20, 36 (evidenced in blue and purple); further rising the threshold to $r_{th}=0.6$, the two latter subgroups reveal a core structure defined by the pairs 19, 35 and 20, 36. An analogous result is found for the sensory system, confirming the hierarchical character of the mouse brain modular structure.

\begin{figure}[t!]
\centering
\includegraphics[width=0.55\textwidth]{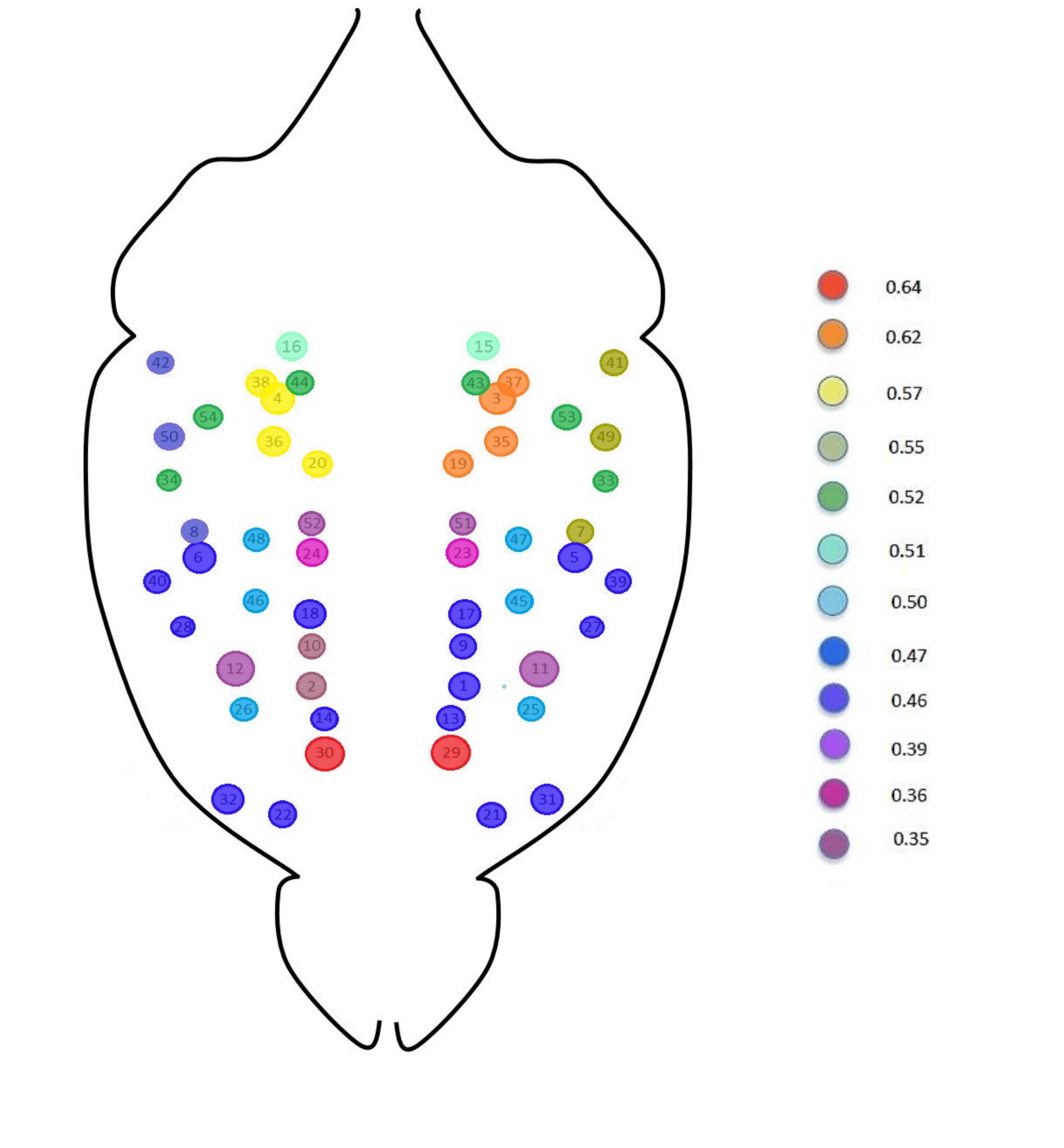}
\caption{Result of the MSF algorithm mapped into the average mouse brain areas. The algorithm works by first sorting the observed correlations in decreasing order and then linking pairs of areas sequentially, with the only limitation that each new link must connect at least one previously disconnected area. Colors correspond to the average correlation value of the links defining each tree composing the forest. See also the map of the neuroanatomical ROI in Appendix.}
\label{fig6}
\end{figure}

Interestingly, the percolation curve for the null model shows a remarkably different trend. Indeed, drawing a matrix (whose distribution of correlations coincides with the observed one for the average mouse) from our ensemble and repeating our percolation analysis leads to a single, sharper transition, with basically no plateaus. This indicates that rising the threshold value leads to the sequential disconnection of individual nodes, which are removed one after the other. This supports the idea that the hierarchical structure observed for the brain connectivity network is genuine, as the stepwise behavior does not emerge in a null model with similar distribution of correlations (the same conclusion holds true for the individual-wise matrices as well).

While this is reassuring, a statistical test is needed to quantify the significance of the experimental trend with respect to the null hypothesis. Our choice of such test moves from the observation that the experimental trend is less steep than the one obtained by running the null model. For this reason, the test statistics we have computed is the steepness of the experimental trend, measured between two points: the pairs $(r'_{th}, 2)$ and $(r''_{th}, 53)$, with $r'_{th}$ and $r''_{th}$ indicating the values of correlations in correspondence of which we detect 2 and 53 communities respectively (we have deliberately excluded the trivial communities represented by the whole brain and the single areas/nodes). The ensemble distribution of our test statistics is shown in fig. \ref{fig4}, together with the experimental point: the latter lies well outside the $95\%$ confidence intervals.
\newline
\newline
\indent On the other hand, as evident upon inspecting fig. \ref{fig3}, classical percolation, in which only the size of the largest component is monitored, detects multiple thresholds in both the experimental network and in the network generated according to our null model. Although this finding provides a significant evidence of the structural differences between the observed average brain and an Erd\"os-Renyi-like graph, for example, it also implies that the claim according to which,  in this ``classical'' version of percolation, revealing multiple thresholds is, by itself, a proof of the hierarchical modular structure of a network is arguable. Indeed, recovering the presence of steps also in the null model seems to suggest that the dynamics of the giant component is (at least) partially encoded into the correlations distributions, while this is no longer true when considering also the remaining components, implying that one of the genuine signatures of the brain self-organization lies in their dynamics.

\paragraph{Minimal Spanning Forest.} 

The MSF algorithm is defined by two simple steps: {\it a} the observed correlations are sorted in reverse order; {\it b)} starting from the largest observed correlation, a link is drawn between the corresponding brain areas. This is done sequentially, with the limitation that any new connection must link at least one previously completely disconnected area.

At each step of the MSF algorithm, either a previously isolated area is assigned to an existing group or two previously isolated areas are linked together. In this way, ``communities'' remain naturally defined by the strength of their internal correlations, while redundant connections are discarded. In particular two different communities are eventually connected by edges whose correlation value is smaller than all the links of both communities. Althought the MSF is not, by itself, a community detetction technique, it provides a means to hierarchically order modules based on the strength of their internal edges. Such modules are tree-shaped and provide information on the structural importance of each area (e.g. its betweeness centrality).

The MSF of average brain is shown in fig. \ref{fig6}. Our analysis reveals that presence of both inter- and intra-hemispheric modules. The module with the strongest internal connectivity is the medial-prefrontal cortex, consistent with the finding that this bilateral structure persists in the percolation analysis at high values of the threshold. The second and third modules in the MSF rank are the right and left hippocampal formation. Interestingly, larger, inter-hemisferic modules, like the one comprising frontal and orbitofrontal cortices, caudate putamen and the amygdala, are characterized by more numerous, but weaker links. Altogether, the MSF structure reflects the hierarchical organization of connectivity modules revealed by our percolation analysis. 

\begin{figure}[t!]
\centering
\includegraphics[width=0.5\textwidth]{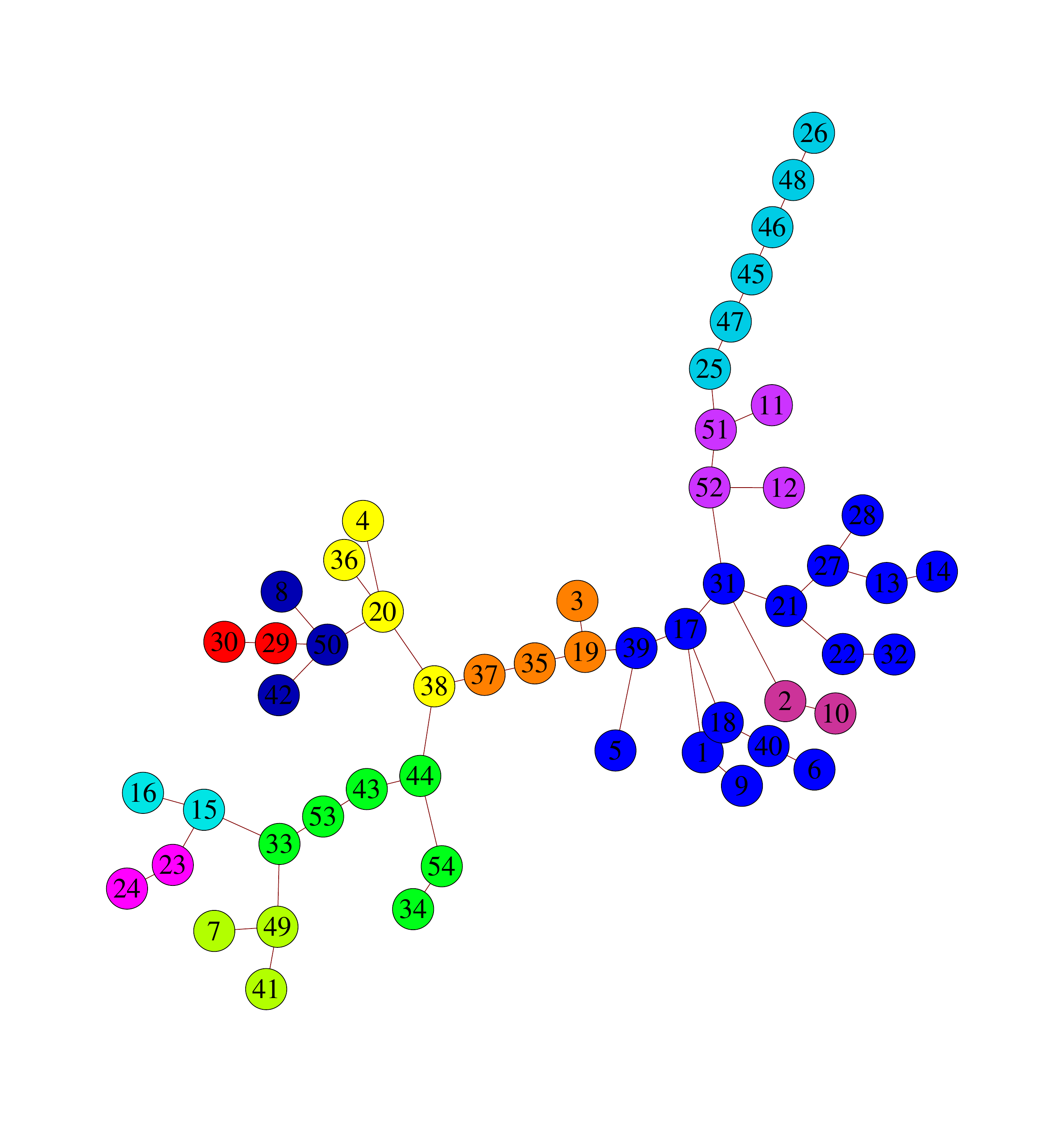}
\caption{MST of our average mouse brain. The MST has been built by connecting the trees of the MSF, with the only limitation that any newly-added link must connect a pair of previously-disconnected trees: a consequence of the MST algorithm is that the correlations within the trees are, on average, higher than the correlations between the trees. The MST also allows us to distinguish between connector and provincial areas. See also the map of the neuroanatomical ROI in Appendix.}
\label{fig7}
\end{figure}

Remarkably, the insular cortex and the secondary somatosensory cortices are found within the same tree, thus showing that the reciprocal structural connectivity among these areas results in a consistent pattern of functional connectivity which has been recently described also using voxelwise community detection approaches \cite{liska}. Similarly, the thalamus is found to be strongly linked to the bed nucleus of stria terminals, consistent with the reciprocal neuroanatomical links connecting these regions \cite{choi}. Interestingly, our MSF reveals a strong functional connection between the visual cortex and the retrosplenial cortex (i.e. between areas 43, 44, 53 and 54), an area that has been recognized as fundamental in tasks like orientation, head movement and processing of visual cues \cite{maguire}. As a last example, the MSF suggests  a role for the temporal association cortex (i.e. 49, 50) in the coordination of the sensori stimuli \cite{squire}, receiving inputs from  the auditory and the rhinal corteces (i.e. 7, 8 and 41, 42).
\newline
\newline
\indent Once the MSF has been built, we can use the remaining correlations in the list to build the Minimal Spanning Tree (MST). As for the forest, only one limitation exists: any new added link must connect a pair of previously-disconnected trees (which become part of the same tree afterwards). Naturally, the links \emph{between} trees are weaker than the links \emph{within} trees and the MSF can be recovered upon removing the weakest links.

The information provided by the MSF can be thus complemented by the information provided by the MST, which gives a clear picture of the mouse brain connectivity skeleton. In particular, the structural role of each area becomes evident and a classification of \emph{connector} areas VS \emph{provincial} areas becomes now possible. Among the most prominent examples of the former are the posterio-ventral hippocampus (i.e. 38) whose physical centrality is recovered as a functional centrality, the parietal association cortex (i.e. 33) which connects all the sensory areas (i.e. the rhinal, auditory and visual ones) and the orbitofrontal cortex (i.e.31) whose physical connections are mirrored by a high degree of functional (inter)-connectivity (e.g. it connects the thalamus and the frontal association cortex).

\section{Conclusions}

In this paper we have presented the results of a network theory-based analysis of a large mouse fMRI dataset, aimed at assessing the hierarchical modular structure of resting state functional connectivity networks in this species. In order to overcome the limitations of currently available techniques, we propose a modified percolation analysis that retains the information on all the connected components of a given network. Our variation of the percolation analysis takes into account negative correlations, and does not require the application of a threshold to binarize the connectivity networks.

Our technique, straightforwardly applicable to experimental correlation matrices, reveals a hierarchically organized modular structure that does not appear in a null model defined by constraining the distribution of the observed correlations. Notably, conventional percolation analysis shows the presence of multiple percolation thresholds also in the null model, thus suggesting that results based on the giant connected component alone maybe misleading.

Our percolation analysis represents a generalization of the classical one. Indeed, while each step of the classical percolation analysis is always mappable into a step of our method, the reverse is not true, since the detection of a newly disconnected module from a secondary component would be completely missed.

We have also computed the Minimal Spanning Forest (MSF) and the Minimal Spanning Tree (MST) for our population-wise mouse brain. The latter represents a faster alternative to the usual community detection techniques, since it identifies modules on the basis of the strengths of their internal correlations. The MSF reveals both intra- and inter-hemispherical modules, and the presence of small, tightly coupled modules alongside with larger subnetworks characterized by weaker internal links. The MST, on the other hand, enables the classification of connector and provincial areas.

Our results indicate that the tools provided by network theory indeed provide additional, non-trivial information on the topology of functional connectivity networks from the mouse brain. This work can be straightforwardly extended to the study of the human brain.

\section*{Appendix}

\subsection*{\it ROI - Regions of interest}

\begin{figure}[t!]
\centering
\includegraphics[width=0.5\textwidth]{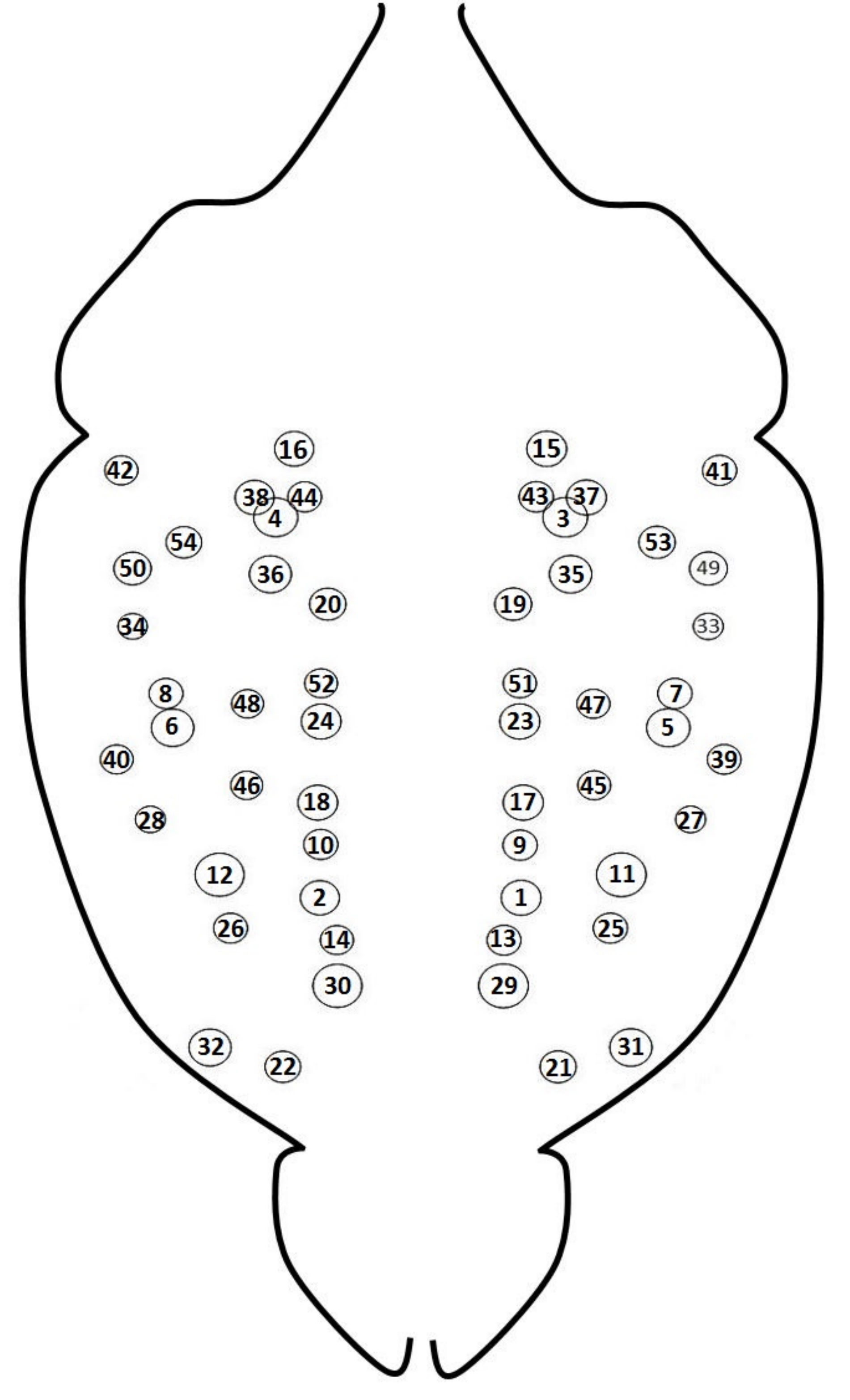}
\caption{The neuroanatomical ROI considered for our analysis, mapped into a mouse brain.}
\label{figA1}
\end{figure}

The list of the neuroanatomical ROI considered for our analysis, together with their abbreviation, is the following (alphabetical order). Fig. \ref{figA1} shows the ROI mapped into a mouse brain.

\begin{enumerate}
\item Acb: accumbens nucleus\_dx;
\item Acb: accumbens nucleus\_sx;
\item AdHC: anterio-dorsal hippocampus\_dx;
\item AdHC: anterio-dorsal hippocampus\_sx;
\item Amy: amygdala\_dx;
\item Amy: amygdala\_sx;
\item Au: auditory cortex\_dx;
\item Au: auditory cortex\_sx;
\item BF: basal forebrain\_dx;
\item BF: basal forebrain\_sx;
\item BNST: bed nucleus of stria terminals\_dx;
\item BNST: bed nucleus of stria terminals\_sx;
\item Cg: cingulate cortex\_dx;
\item Cg: cingulate cortex\_sx;
\item Collicoli: collicoli\_dx;
\item Collicoli: collicoli\_sx;
\item Cpu: caudate putamen\_dx;
\item Cpu: caudate putamen\_sx;
\item DG: dentate gyrus\_dx;
\item DG: dentate gyrus\_sx;
\item FrA: frontal association cortex\_dx;
\item FrA: frontal association cortex\_sx;
\item Hypo: hypothalamus\_dx;
\item Hypo: hypothalamus\_sx;
\item Ins: insular cortex\_dx;
\item Ins: insular cortex\_sx;
\item M: motor cortex\_dx;
\item M: motor cortex\_sx;
\item mPFC: medial prefrontal cortex\_dx;
\item mPFC: medial prefrontal cortex\_sx;
\item OFC: orbitofrontal cortex\_dx;
\item OFC: orbitofrontal cortex\_sx;
\item Parietal\_Ass: parietal association cortex\_dx;
\item Parietal\_Ass: parietal association cortex\_sx;
\item pDG: posterior dentate gyrus\_dx;
\item pDG: posterior dentate gyrus\_sx;
\item pHC: posterio-ventral hippocampus\_dx;
\item pHC: posterio-ventral hippocampus\_sx;
\item Pir: piriform cortex\_dx;
\item Pir: piriform cortex\_sx;
\item Rhinal: rhinal cortex\_dx;
\item Rhinal: rhinal cortex\_sx;
\item RS: retrosplenial cortex\_dx;
\item RS: retrosplenial cortex\_sx;
\item S1: primary somatosensory cortex\_dx;
\item S1: primary somatosensory cortex\_sx;
\item S2: secondary somatosensory cortex\_dx;
\item S2: secondary somatosensory cortex\_sx;
\item TeA: temporal association cortex\_dx;
\item TeA: temporal association cortex\_sx;
\item Th: thalamus\_dx;
\item Th: thalamus\_sx;
\item Vctx: visual cortex\_dx;
\item Vctx: visual cortex\_sx.
\end{enumerate}

\subsection*{\it From time series to correlation matrices}

\begin{figure}[t!]
\centering
\includegraphics[width=0.5\textwidth]{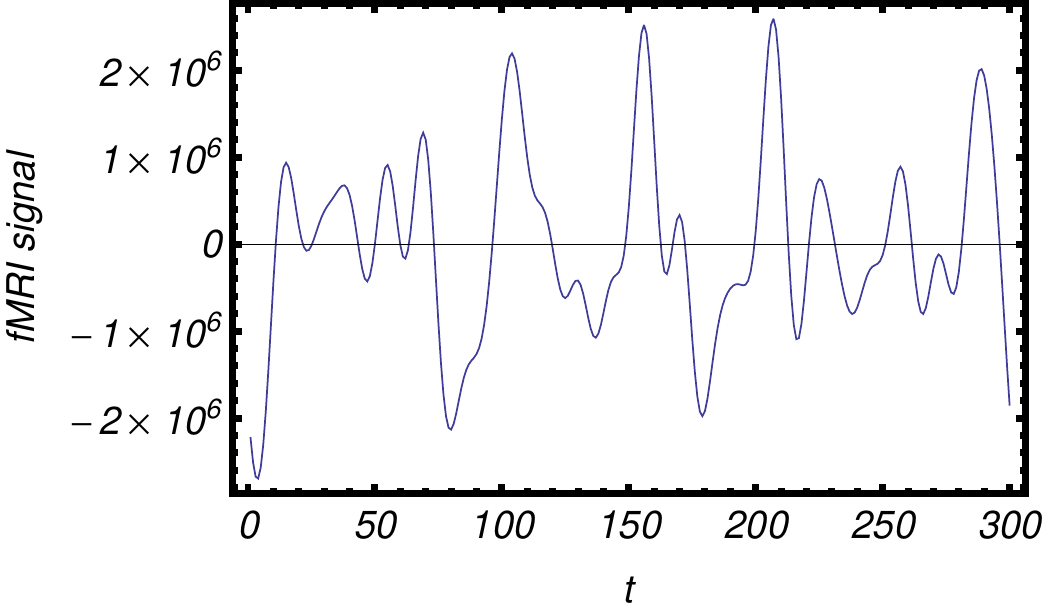}
\includegraphics[width=0.5\textwidth]{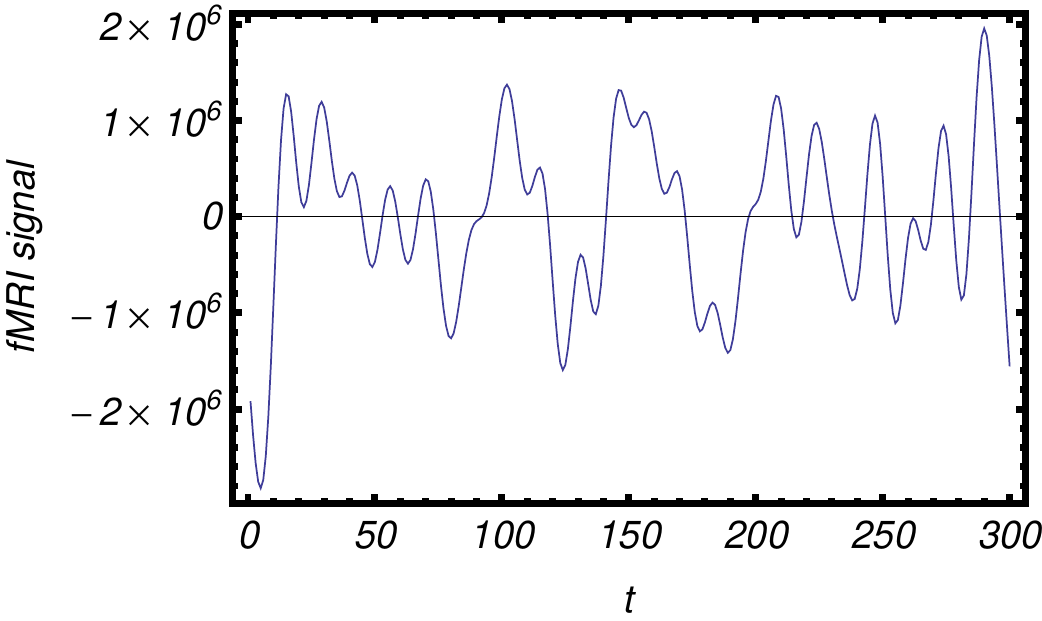}
\caption{fMRI BOLD-signals corresponding to the right cingulate cortex (top) and left cingulate cortex (bottom) of the brain BE\_ag130207a.}
\label{figA2}
\end{figure}

Our data consist of 41 sets of 54 fMRI BOLD-signals each, collected as the time series shown in fig. \ref{figA2}.

The information carried by each mouse-specific set of time series has been condensed into a correlation matrix, whose generic entry $C_{ij}$ is the Pearson coefficient between time series $X^i$ and $X^j$, defined as

\begin{eqnarray}
C_{ij}&=&\frac{\mbox{Cov}[X^i, X^j]}{\sqrt{\mbox{Var}[X^i]\cdot\mbox{Var}[X^j]}}=\nonumber\\
&=&\frac{\sum_{t=1}^T(X_t^i-m^i)(X^j_t-m^j)}{\sqrt{\sum_{t=1}^T(X^i_t-m^i)^2\cdot\sum_{t=1}^T(X^j_t-m^j)^2}}
\end{eqnarray}

\noindent where $m^i=\frac{\sum_{t=1}^TX^i_t}{T}$ and $T$ is the total temporal length of the series.
\newline
\newline
\indent In order to create an average adjacency matrix describing brain functional connectivity at the population level, subject-wise matrices were first Fisher-transformed, i.e.

\begin{equation}
z_{ij}=\frac{1}{2}\ln\left(\frac{1+C_{ij}}{1-C_{ij}}\right)=\mbox{arctanh}(C_{ij}),
\end{equation}

\noindent then averaged across subjects 

\begin{equation}
\overline{z}_{ij}=\frac{\sum_{n=1}^{54}z_{ij}^n}{54},\:\forall\:i,j
\end{equation}

\noindent (i.e. the generic entry of the average Fisher-transformed matrix is the arithmetic mean of the corresponding individual entries, $z^1_{ij}$, $z^2_{ij}\dots z^{54}_{ij}$) and then back-transformed:

\begin{equation}
\overline{C}_{ij}=\mbox{tanh}(\overline{z}_{ij}),\:\forall\:i,j.
\end{equation}

\begin{acknowledgements}
We acknowledge support from the EU project FET-Open FOC (grant num. 255987), the FET project SIMPOL (grant num. 610704) and the FET project DOLFINS (grant num. 640772).
\end{acknowledgements}

\end{document}